\documentclass[aps,prb,twocolumn,floatfix,showpacs,superscriptaddress]{revtex4}

\usepackage{graphicx}
\usepackage{bbm}
\usepackage{amsmath,amssymb}

\newcommand{\be}{\begin{equation}}
\newcommand{\ee}{\end{equation}}
\newcommand{\bea}{\begin{eqnarray}}
\newcommand{\eea}{\end{eqnarray}}

\begin{document}
\title{Optical control of entangled states in quantum wells}
\author{E. R{\"a}s{\"a}nen}
\email[Electronic address:\;]{esa.rasanen@tut.fi}
\affiliation{Department of Physics, Tampere University of Technology, 
FI-33101 Tampere, Finland}
\affiliation{Nanoscience Center, Department of Physics, University of
  Jyv\"askyl\"a, FI-40014 Jyv\"askyl\"a, Finland}
\affiliation{Physics Department, Harvard University, Cambridge, Massachusetts 02138, USA}
\author{T. Blasi}
\affiliation{Physics Department, Harvard University, Cambridge, Massachusetts 02138, USA}
\affiliation{Physik Department, Technische Universit\"at M\"unchen, D-85747 Garching, Germany}
\author{M. F. Borunda}
\affiliation{Physics Department, Harvard University, Cambridge, Massachusetts 02138, USA}
\affiliation{Department of Physics, Oklahoma State University, Stillwater, Oklahoma 74078, USA}
\author{E. J. Heller}
\affiliation{Physics Department, Harvard University, Cambridge, Massachusetts 02138, USA}
\affiliation{Department of Chemistry and Chemical Biology, Harvard University, Cambridge, Massachusetts 02138, USA}
\date{\today}

\begin{abstract}
We present theory and calculations for coherent 
high-fidelity quantum control of many-particle states 
in semiconductor quantum wells. We show that coupling
a two-electron double quantum dot to a terahertz optical source 
enables targeted excitations that are one to two orders of magnitude 
faster and significantly more accurate than those obtained 
with electric gates. The optical fields subject to 
physical constraints are obtained through quantum 
optimal control theory that we apply in conjunction with the 
numerically exact solution of the time-dependent Schr\"odinger 
equation. Our ability to coherently control arbitrary 
two-electron states, and to maximize the entanglement,
opens up further perspectives in solid-state quantum 
information.
\end{abstract}

 
\maketitle


Quest for solid-state quantum computing~\cite{loss,bitreview} 
in low-dimensional semiconductor materials has led
to experimental breakthroughs in the initialization, 
processing, and readout of single- and two-particle
states in coupled quantum 
dots.~\cite{petta,conditionaldynamics,lieven,science2012}
The majority of proposed quantum computation schemes 
focus on spin control~\cite{spinreview} due to the long
decoherence times up to microseconds. However, rapid
development of light sources and ultrafast science
 in atomic physics~\cite{attoreview} is likely to bring 
methods and techniques to coherent control of
electric charge~\cite{hayashi} in low-dimensional systems. 

Electron dynamics in double quantum dots (DQDs) has 
been subject to various theoretical approaches.  
Beyond few-level models, realistic DQDs have 
been controlled with 
gate voltages~\cite{murgida1,murgida2,kataoka} and optimized 
laser pulses.~\cite{2D_DQD,1D_DQD,salen,nepstad}
Murgida and co-workers~\cite{murgida1,murgida2} 
presented an scheme to reach desired excited 
states in two-electron DQDs by ``navigating'' in the energy 
spectrum with electric fields -- an approach widely used
for two- and three-level systems.~\cite{sangouard}
Here we focus on a similar
quasi-one-dimensional setup and show that optical fields
can lead to almost $100\%$ fidelities of targeted 
entangled two-electron states in tens of picoseconds,
thus improving {\em both} the operation time and the fidelity
from earlier works by 1-2 orders of magnitude.
In this context, we extend the previous 
single-particle~\cite{2D_DQD,1D_DQD} and 
two-particle~\cite{salen,nepstad} optimization studies
into an all-round scheme to coherently control an arbitrary 
transition in a two-electron DQD. Moreover, we show
that the control procedure validates for the maximization
of the entanglement itself, which might have further
implications for quantum information in solid-state devices.

\begin{figure}
\includegraphics[width=0.75\columnwidth]{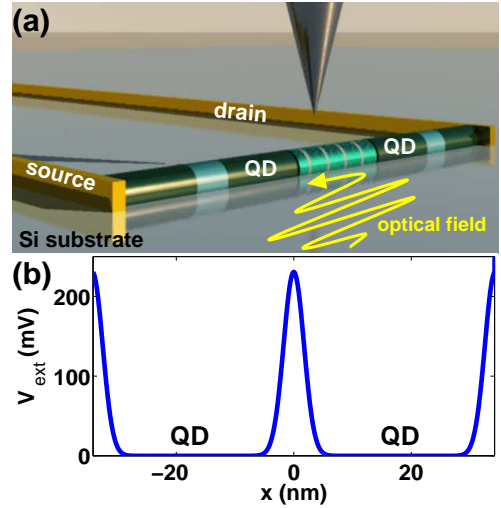}
\caption{(color online) (a) Schematic picture of
the proposed setup. Two quantum dots in a 
quasi-one-dimensional nanowire are separated by
tunnel barriers, and the central one can be adjusted
by a conducting tip. (b) One-dimensional model potential 
for a double quantum dot used in the calculations.
}
\label{fig1}
\end{figure}

Our model system is illustrated in Fig.~\ref{fig1}(a).
We consider a quasi-one-dimensional nanowire fabricated, 
e.g., by epitaxial methods.~\cite{jensen}
Two QDs are separated by tunnel barriers, and the
strength and position of the central barrier can
be adjusted by a scanning gate microscope tip.~\cite{boyd,bles} 
Here we assume the tip to be located in the middle, so that
the one-dimensional model potential has the shape shown
in Fig.~\ref{fig1}(b). The edges of the barriers are
softened by Gaussian functions. 

For general applicability of our results,
and to explicitly compare with 
Refs.~\onlinecite{murgida1} and \onlinecite{murgida2}, we 
consider GaAs material
parameters with the effective mass $m^*=0.067\,m_e$ and
dielectric constant $\epsilon=12.4\,\epsilon_0$. 
We use effective atomic units (a.u.) throughout the
paper apart from the physical model in Fig.~\ref{fig1}(b).
The energies, lengths, and times scale as 
$E_{h}^\ast=(m^\ast/m_0)/(\varepsilon/\varepsilon_0)^2E_h\approx 11\,\mathrm{meV}$,
$a_0^\ast=(\varepsilon/\varepsilon_0)/(m^\ast/m_0) a_0\approx 10\,\mathrm{nm}$, and
$t_0^\ast=\hbar/E_h^\ast\approx 55\,\mathrm{fs}$, respectively.

The two-electron problem in one dimension is solved numerically
exactly by considering electron coordinates $x_1$ and $x_2$ and 
a soft-Coulomb interaction of the form 
$V_{\rm int}(x_1,x_2)=[(x_1-x_2)^2+\alpha]^{-1/2}$ with a softening parameter 
$\alpha=0.25$.
The Hamiltonian reads
$H = K + V_{ext}(x_1,x_2) + V_{int}(x_1, x_2)- \mu \epsilon(t)$, where
$K$ is the kinetic-energy operator and 
$V_{\rm ext}(x_1,x_2) = {\max}[V(x_1),V(x_2)]$ with $V(x_i)$ is the
external confinement potential for each individual electron shown
in Fig.~\ref{fig1}(b). It represents a double quantum well with 
a depth 220 meV, width 35 nm, and interwell barrier 5 nm.
In the Hamiltonian $\hat{\mu}=-(x_1,x_2)$ is the dipole operator and
$\boldsymbol{\epsilon}(t)$ is the optical field.
We point out that the spatial wave functions across the
diagonal $x_1=x_2$ are symmetric, since we consider only antisymmetric
two-electron singlet states (so that the total wave function
is antisymmetric). We could focus on triplets in a similar manner
by considering only spatially antisymmetric wave functions.
A small distortion potential 
$V_{d}(x_1,x_2)=0.001(x_1+x_2)$, corresponding to a small static field, is
added to the Hamiltonian to distinguish degenerate eigenstates
as in a real setup, where impurities are always present.

Our task is to {\em optimize} $\boldsymbol{\epsilon}(t)$
such that the two-electron wave function $\left.|\Psi(x_1,x_2)\right>$
reaches the desired target state $\left.|\Phi_{\rm F}\right>$ at 
a fixed time $T$ by maximizing the overlap 
$\left|\big<\Psi(x_1,x_2,T)|\Phi_{\rm F}\big>\right|^{2}$.
To this end, we apply quantum optimal control 
theory~\cite{oct} (OCT) as explained in 
Ref.~\onlinecite{1D_DQD}. This gives the optimized field
$\boldsymbol{\epsilon}(t)$ through an iterative 
procedure of the time-dependent Schr\"odinger equation.
As physical constraints, we set the maximum allowed frequency
in the field to $\omega_{\rm max}$ and the
fluence $F$, i.e., the integrated intensity 
(values given below). We use the OCT scheme
of Werschnik and Gross~\cite{Werschnik2} and 
apply the {\tt octopus} code~\cite{octopus}
in real space and real time.


\begin{figure}
\includegraphics[width=0.85\columnwidth]{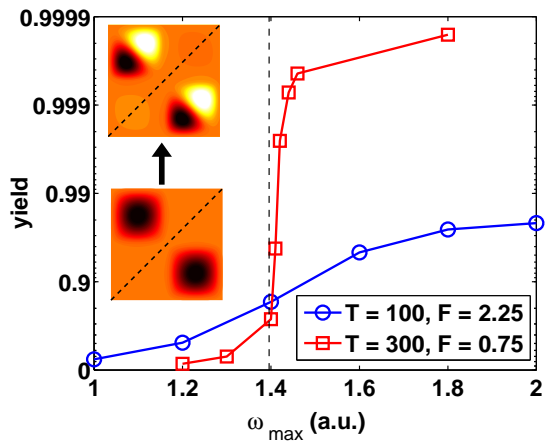}
\caption{(color online) Yield as a function of
the maximum frequency in the optimized pulse
to excite the system from the ground state to
the fourth state. The electrons remain delocalized but
a node is created in the wells (see the inset).
Two different pulse lengths and fluences are considered
(see text). 
}
\label{fig2}
\end{figure}

First we consider an excitation where the electrons stay
in separate dots (we call this a delocalized configuration)
but a node is created in both dots as shown in the
inset of Fig.~\ref{fig2}. This corresponds to an
excitation from the ground state $\left.|1\right>$ to the fourth
state $\left.|4\right>$ ($\left.|2\right>$ and $\left.|3\right>$ 
are localized configurations with both electrons being 
in the left and right well, respectively)
The obtained yield strongly
depends on $\omega_{\rm max}$ allowed in the optimization,
and with a pulse length $T=300$ (17 ps) we find a strong
increase slightly above $\omega_{\rm max}= 1.4$. This is
simply due to the resonance at $E_4-E_1=1.3957$
(dashed line), which is a crucial frequency to be included
in the optimization to reach extremely high yields.
A shorter ($T=100$) but more intensive pulse lacks this
behavior due to the tunneling character of the excitation;
the Keldysh parameter -- used in atoms to assess whether the process
is of tunneling or (multi)photon character --
is inversely proportional to the peak 
intensity.~\cite{epl} However, the shorter pulse leads to
higher yields than the longer (and less intensive) one
with filter frequencies smaller than the resonance frequency.
To underline the effectiveness of OCT in Fig.~\ref{fig2} 
it is noteworthy that a {\em non-optimized}
pulse with the resonant frequency and $T=100$ and $T=300$,
leads, {\em at most}, to yields $\sim 70\%$ and $\sim 95\%$,
respectively, whereas the optimized pulses, with 
$\omega_{\rm max}$ about $10\%$ above the resonance, lead 
to respective yields of $\sim 95\%$ and $\sim 99.95\%$.

\begin{figure}
\includegraphics[width=0.95\columnwidth]{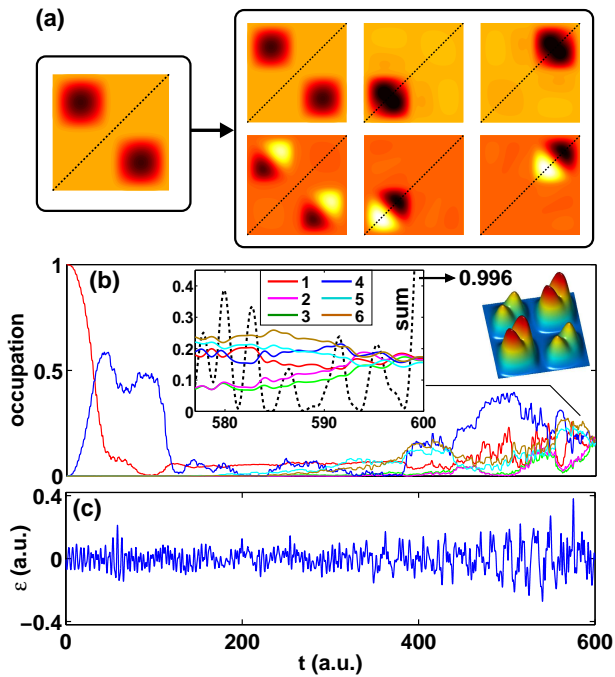}
\caption{(color online) (a) Excitation process
from the ground state to an equal superposition of 
six eigenstates. (b) Occupations of the states during
the optimized process. At the end (inset) the states
reach $\sim 1/6$ occupations while a total yield
of  $99.6\%$ is obtained for the superposition 
(total density shown in the right inset). (c) Optimized
pulse analyzed in Fig.~\ref{fig4}.
}
\label{fig3}
\end{figure}

In Fig.~\ref{fig3} we show our results for an optimized 
excitation from the ground state $\left.|1\right>$ to a superposition 
that consists of the six lowest states, i.e., the target
state can be written as
$\left.|\Phi_{\rm F}\right>=(1/\sqrt{6})\left(\left.|1\right>+\left.|2\right>+\left.|3\right>+\left.|4\right>+\left.|5\right>+\left.|6\right>\right)$
[Fig.~\ref{fig3}(a)]. The pulse length and fluence are fixed to 
$T=600$ ($=33$ ps) and $F=6$ ($=71$ meV), respectively, and the 
filter frequency is $\omega_F=3$ ($=8.6$ THz). As shown in 
Fig.~\ref{fig3}(b) we find an extremely high total yield $99.6\%$,
and the occupations of the individual states converge close to $1/6$
(see the inset). The final density, consisting of both localized and
delocalized contributions, is shown in the right inset. 

It was shown in Ref.~\onlinecite{murgida2} that properly adjusted electric
fields can drive a similar system to a three-state superposition
in 1000 ps with $96.56\%$ final overlap. In this respect, our
optical scheme appears to be about 30 times faster and
an order of magnitude more accurate, even though the target system
is more complex (six states instead of three). The price to pay is 
the rather complicated control field in Fig.~\ref{fig3}(c)
that, however, satisfies realistic constraits and, secondly,
can be thoroughly understood with the following
analysis. It is important to note that the optimized field as
the one in Fig.~\ref{fig3}(c)
corresponds to a {\em local} maximum in a search space, i.e., there
might be other solutions yielding similar, or even better outcomes.

\begin{figure}
\includegraphics[width=0.95\columnwidth]{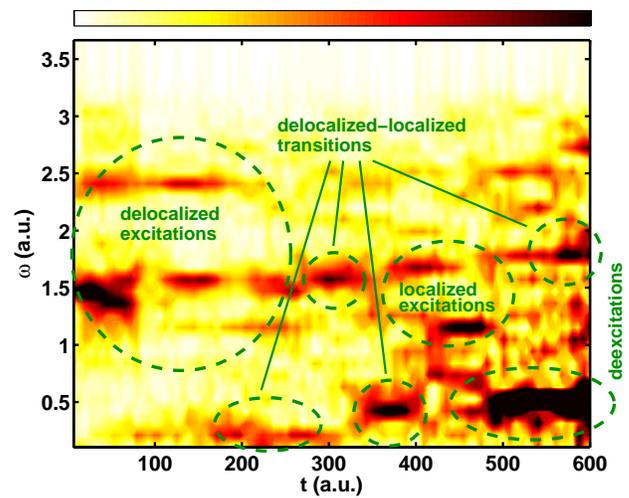}
\caption{(color online) Time-frequency map of the optimized field
shown in Fig.~\ref{fig3}(c). Amplitudes of the appearing frequencies
as a function of time are shown in colorscale 
(white: small, black: large). Several (de)excitation processes can be 
distinguished (see text).
}
\label{fig4}
\end{figure}

Figure~\ref{fig4} shows a time-frequency map 
of the optimized pulse in Fig.~\ref{fig3}(c). 
The amplitudes 
of the appearing frequencies ({\em y} axis) during the 
pulse ({\em x} axis) are shown in colorscale. Time
frames of $\Delta t=T/10=60$ were used in the Fourier 
transforms. Several distinctive features can be seen
in the map, and they can be associated 
with specific (de)excitation processes. First, in the beginning of the
pulse we find pronounced frequencies around $\omega\sim 1.5$ 
and $2.4$ corresponding to {\em delocalized} excitations, i.e.,
nodes are created in left-right configurations,
with $\omega_{1,4}=E_4-E_1=1.40$ (as in Fig.~\ref{fig2}), 
$\omega_{4,7}=1.43$, and $\omega_{4,10}=2.30$. 
Around $t=400\ldots 500$ we find similar processes for
{\em localized} configurations (left-left or right-right)
with $\omega_{2,5}=\omega_{3,6}=1.57$ and 
$\omega_{5,8}=\omega_{6,9}=1.06$. In between, we find
contributions corresponding to {\em transitions} between
delocalized and localized configurations. They occur
through superpositions of the localized states and
require frequencies equal to energy gaps between delocalized
and localized states. Finally, close to the end of the
pulse we find a very large amplitude at $\omega~\sim 0.5$
alongside several small-scale features.
They correspond to several {\em deexcitations} from 
high-lying occupations back to the target eigenstates as discussed 
within Fig.~\ref{fig3}. Overall, with the time-frequency map we
are able to understand most features of the optimization
process despite the complicated shape of the pulse in
the time domain [Fig.~\ref{fig3}(c)]. On the basis of Fig.~\ref{fig2} 
it can be expected that filtering out the dominating frequencies 
in Fig.~\ref{fig4} may significantly lower the yield. 
Quantitative assessment of 
the sensitivity of the yield to different frequencies is 
the subject of future work.

Finally, we examine how the correlation entropy and 
entanglement are evolved during the controlled excitation
to a superposition. 
The correlation entropy~\cite{ref1,ref2} is defined as
$S=-\sum_k^N n_k\ln n_k$, where the sum runs over the
$N$ lowest eigenvalues $n_k$ of the one-body density matrix. 
The maximum value for $S$ is obtained when $n_k=1/N$, i.e., 
when all the states are equally occupied. 

The entanglement, on the other hand,
is defined here in terms of the von Neumann entropy~\cite{ref3}
$E_j=-\mbox{Tr}\left(\rho_j\log_2\rho_j\right)$, 
where $\rho_j=\mbox{Tr}_j\left(\left| \Psi \right\rangle \left\langle \Psi\right| \right)$
is the reduced density matrix, and the trace is taken
over the $j$:th site (1 or 2) of the DQD.
The three possible well occupations 
are $\left|q\right\rangle_j$ with $q=0,1,2$ being the charge
of the well. Now the reduced density matrix becomes
$\rho_j=p_{0,j}\left|0\right\rangle_j\langle 0|_j+p_{1,j}\left|1\right\rangle_j\langle 1|_j+p_{2,j}\left|2\right\rangle_j\langle 2|_j$ (occupations $\left|1\right\rangle_j$ and 
$\left|2\right\rangle_j$ should not be
mixed with {\em states} $\left|1\right\rangle$ and $\left|2\right\rangle$).
Here $p_{i,j}$ is the probability of the $j$th site being occupied 
with $i$ electrons. Therefore, the entanglement can be written as
$E_j=\sum_{i=0}^{2}p_{i,j}\log_2p_{i,j}$.

\begin{figure}
\includegraphics[width=0.85\columnwidth]{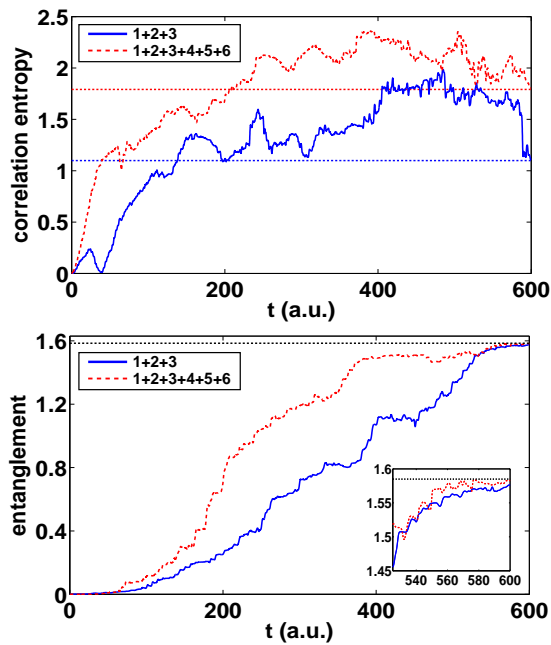}
\caption{(color online) Correlation entropy (upper panel)
and the entanglement (lower panel) in excitation processes
from the ground state the superposition of the first
three states (solid lines) and to the superposition of the
first six states (dashed lines).
The latter process is the same as in Figs.~\ref{fig3} and \ref{fig4}.
The horizontal dotted lines in (a) and (b) mark the ideal and
maximum values, respectively (see text).
}
\label{fig5}
\end{figure}

In Fig.~\ref{fig5} we plot the correlation entropy (upper panel)
and the entanglement (lower panel) for two excitations:
from the ground state to the superposition of the 
first three states, i.e., $\left.|\Phi_{\rm F}\right>=(1/\sqrt{3})\left(\left.|1\right>+\left.|2\right>+\left.|3\right>\right)$,
and to the superposition to the first six states
(the same as in Figs.~\ref{fig3} and \ref{fig4}).
The correlation entropy oscillates strongly in both cases but
eventually reaches the values $1.1112$ and $1.8062$, respectively, that are 
very close to the ideal values $\ln(3)\approx 1.0986$ and
$\ln(6)\approx 1.7918$ (dotted lines). 
This means that at $t=T$ the eigenstates in the
target superpositions are almost equally occupied. The higher 
entropies at $t<T$ are due to the distribution of the electron 
occupations to a large number of states -- about $20$ states
are occupied by more than $1\%$ in the process. 

The entanglement in the lower panel of Fig.~\ref{fig5} approaches 
the theoretical maximum value $\log_2(3)\approx 1.5850$ (dotted line).
For the three-state and six-state optimization processes we find
$1.5764$ and $1.5830$, respectively. This can be associated
with the {\em maximization of the entanglement} with $99.5\%$ and 
$99.9\%$ fidelities, that are comparable to the respective target-state
overlaps of $99.5\%$ and $99.6\%$. Thus, through the proposed setup we are
able to find maximally entangled two-electron states in
a targeted fashion. We also note that the entanglement could be 
explicitly measured in the DQD by detecting the charge~\cite{ref4} 
on one of the 
wells leaving the other well unaffected. This procedure would directly 
correspond to tracing out one site of the DQD and reducing the density 
matrix of the whole system to 
$\rho_j=\mbox{Tr}_j\left(\left| \Psi \right\rangle \left\langle \Psi\right| \right)$.

We point out three aspects in the practical realization of the proposed
control scheme. First, the actual confining potential and thus the
Hamiltonian could be determined through a supporting experiment, e.g.,
by measuring the single-particle spectrum and constructing $V_{\rm ext}$ 
through inversion.~\cite{jens} The optimization could then be performed
within this specific $V_{\rm ext}$. Secondly, it is important to note that 
the physical constraints of the optical pulses required to induce the 
proposed transitions are within reach of the present experimental 
capabilities. Terahertz frequencies can be routinely produced with, 
e.g., quantum cascade lasers and they can be shaped even in the 
femtosecond time scale.~\cite{book} For example, in a recent 
experiment~\cite{wirth} femtosecond pulses were created in a synthesized 
fashion by separately controlling the infrared, visible, and ultraviolet 
contributions, including their chirp, carrier envelope phase, delay, and
energy (beam size). Therefore, coherent control with THz optics 
is likely to become a feasible scheme in solid-state quantum
information. On the other hand, voltage control with gates 
is so far strictly limited to GHz time scales. Finally, as the third
critical aspect we mention coherence that highly depends
on the external conditions in a particular device. We believe that
the time scales presented here are coherently reachable, although
we cannot rule out the possiblity of decoherence due to hyperfine effects 
or/and interactions with optical and acoustic phonons.


To summarize, we have shown through numerically exact calculations
that coupling a two-electron  double quantum dot to an optimized 
optical source enables coherent excitations
to entangled states with an extremely high fidelity. 
In the proposed setup the pulses produced by quantum optimal 
control theory can be exposed to realistic constraints without 
losing the accuracy. Time-frequency analysis gives invaluable 
insight on the physical processes in the evolution towards
the desired superposition. We show that entanglement of
delocalized and localized electron configurations can
be reached up to $99.9\%$ of the theoretical maximum. 
We hope that the proposed scheme motivates further experiments 
in the initialization, control, and readout of electric charge in 
solid-state devices.

\begin{acknowledgments}
We thank Robert Westervelt for helpful discussions.
M.F.B. and E.J.H. were supported by the Department of 
Energy, office of basic science (grant DE-FG02-08ER46513),
E.R. by the Academy of Finland, 
the Wihuri Foundation, and the Magnus Ehrnrooth 
Foundation, and T.B. by Studienstiftung des deutschen 
Volkes. CSC Scientific Computing Ltd. in Finland 
and the Odyssey cluster supported by the FAS Science Division 
Research Computing Group at Harvard University
are acknowledged for computational resources.
\end{acknowledgments}


\begin{thebibliography}{ll}

\bibitem{loss} D. Loss and D. P. DiVincenzo, Phys. Rev. A {\bf 57}, 120 (1998).

\bibitem{bitreview} For a perspective, see D. P. DiVincenzo, Science {\bf 309}, 2173 (2005).


\bibitem{petta} J. R. Petta, A. C. Johnson, J. M. Taylor, E. A. Laird, A. Yacoby, M. D. Lukin, C. M. Marcus, 
M. P. Hanson, and A. C. Gossard, Science {\bf 309}, 2180 (2005).

\bibitem{conditionaldynamics} L. Robledo, J. Elzerman, G. Jundt, 
M. Atat\"ure, A. H\"ogele, S. F\"alt, and A. Imamoglu, Science {\bf 320}, 772 (2008).

\bibitem{lieven} K. C. Nowack, M. Shafiei, M. Laforest, G. E. D. K. Prawiroatmodjo, L. R. Schreiber, C. Reichl, W. Wegscheider, L. M. K. Vandersypen, Science {\bf 333}, 1269 (2011).

\bibitem{science2012} M. D. Shulman, O. E. Dial, S. P. Harvey, 
H. Bluhm, V. Umansky, and A. Yacoby, Science {\bf 336}, 202 (2012).



\bibitem{spinreview} For a review, see R. Hanson, L. P. Kouwenhoven, J. R. Petta, S. Tarucha,
L. M. K. Vandersypen, Rev. Mod. Phys. {\bf 79}, 1217 (2007).

\bibitem{attoreview} 
L. Gallmann, C. Cirelli and U. Keller, Ann. Rev. Phys. Chem. {\bf 63}, 447 (2012).

\bibitem{hayashi} T. Hayashi, T. Fujisawa, H. D. Cheong, Y. H. Jeong, 
and Y. Hirayama, Phys. Rev. Lett. {\bf 91}, 226804 (2003).

\bibitem{murgida1} G. E. Murgida, D. A. Wisniacki, and P. I. Tamborenea,
Phys. Rev. Lett. {\bf 99}, 036806 (2007).

\bibitem{murgida2}G. E. Murgida, D. A. Wisniacki, and P. I. Tamborenea,
Phys. Rev. B {\bf 79}, 035326 (2009).

\bibitem{sangouard} N. Sangouard, S. Guerin, M.
Amniat-Talab, and H. R. Jauslin, Phys. Rev. Lett. {\bf 93}, 223602 (2004).

\bibitem{kataoka} M. Kataoka, M. R. Astley, A. L. Thorn, D. K. L. Oi, C. H. W. Barnes, C. J. B. Ford, D. Anderson, G. A. C. Jones, I. Farrer, D. A. Ritchie, and M. Pepper, Phys. Rev. Lett. {\bf 102}, 156801 (2009).

\bibitem{2D_DQD} E. R{\"a}s{\"a}nen, A. Castro, J. Werschnik, A. Rubio, and E. K. U. Gross,
Phys. Rev. B {\bf 77}, 085324 (2008).

\bibitem{1D_DQD} A. Putaja and E. R\"as\"anen, 
Phys. Rev. B {\bf 82}, 165336 (2010).

\bibitem{salen} L. S\ae{}len, R. Nepstad, I. Degani, and J. P. Hansen, Phys. Rev. Lett. {\bf 100}, 046805 (2008).

\bibitem{nepstad} R. Nepstad, L. S\ae{}len, I. Degani, and J. P. Hansen, 
J. Phys.: Condens. Mat. {\bf 21}, 215501 (2009).

\bibitem{jensen} L. E. Jensen, M. T. Björk, S. Jeppesen, A. I. Persson, 
B. J. Ohlsson, and L. Samuelson, Nano Lett. {\bf 4}, 1961 (2004).

\bibitem{boyd} E. E. Boyd, K. Storm, L. Samuelson, and
R. M. Westervelt, Nanotechnology {\bf 22}, 185201 (2011).

\bibitem{bles} A. C. Bleszynski, F. A. Zwanenburg, R. M. Westervelt, A. L. Roest, E. P. A. M. Bakkers, and . P. Kouwenhoven, Nano Lett. {\bf 7}, 2559 (2007).

\bibitem{oct} A. P. Peirce, M. A. Dahleh, and H. Rabitz, 
Phys. Rev. A {\bf 37}, 4950 (1988); R. Kosloff, S. A. Rice, P. Gaspard, S. Tersigni, and D. J. Tannor,
Chem. Phys. {\bf 139}, 201 (1989).




\bibitem{Werschnik2} J. Werschnik and E. K. U. Gross, J. Opt. B: Quantum Semiclass. Opt. {\bf 7}, S300 (2005).

\bibitem{octopus} 
A. Castro, H. Appel, M. Oliveira, C. A. Rozzi, X. Andrade,
F. Lorenzen, M. A. L. Marques, E. K. U. Gross, and A. Rubio,
Phys. Stat. Sol. (b) {\bf 243}, 2465 (2006).


\bibitem{epl}  A. Castro, E. R\"as\"anen, A. Rubio, and 
E. K. U. Gross, Europhys. Lett. {\bf 87}, 53001 (2009).

\bibitem{ref1} P. Ziesche, O. Gunnarsson, W. John, and H. Beck,
Phys. Rev. B {\bf 55}, 10270 (1997).
 
\bibitem{ref2} Z. Huang, H. Wang, and S. Kais,
J. Mod. Opt. {\bf 53}, 2543 (2006).

\bibitem{ref3} P. Zanardi
Phys. Rev. A {\bf 65}, 042101 (2002).

\bibitem{ref4} M. Field, C. G. Smith, M. Pepper, D. A. Ritchie, 
J. E. F. Frost, G. A. C. Jones, and D. G. Hasko,
Phys. Rev. Lett. {\bf 70}, 1311 (1993).

\bibitem{jens} E. R\"as\"anen, J. K\"onemann, 
R. J. Haug, M. J. Puska, and R. M. Nieminen,
Phys. Rev. B {\bf 70}, 115308 (2004).

\bibitem{book} Y.-S. Lee, 
{\em Principles of Terahertz Science and Technology}, Springer, 2009.

\bibitem{wirth} A. Wirth, M. Th. Hassan, I. Grguras, J. Gagnon, A. Moulet, T. T. Luu, S. Pabst, R. Santra, Z. A. Alahmed, A. M. Azzeer, V. S. Yakovlev, V. Pervak, F. Krausz, and E. Goulielmakis, Science {\bf 334}, 195 (2011).


\end{thebibliography}
\end{document}